\documentclass[preprint]{aastex631}
\usepackage[utf8]{inputenc}

\usepackage{graphicx}
\usepackage{xcolor}
\usepackage{booktabs}
\usepackage{color, colortbl}
\definecolor{LightCyan}{rgb}{0.88,1,1}
\usepackage{enumerate}
\revised{December 19, 2024}
\submitjournal{ApJL}
\shorttitle{Trappist1-d}
\shortauthors{Way}

\begin{document}

\title{TRAPPIST-1 d: Exo-Venus, Exo-Earth or Exo-Dead?}

\correspondingauthor{M.J.~Way}
\email{Michael.Way@nasa.gov}

\author[0000-0003-3728-0475]{M.J.~Way}
\affiliation{NASA Goddard Institute for Space Studies, 2880 Broadway, New York, NY 10025, USA}
\affiliation{GSFC Sellers Exoplanet Environments Collaboration, NASA Goddard Space Flight Center, Greenbelt, MD 20771, USA}
\affiliation{Theoretical Astrophysics, Department of Physics and Astronomy, Uppsala University, Uppsala, SE-75120, Sweden}

\date{July 2023}

\begin{abstract}
TRAPPIST-1 d is generally assumed to be at the boundary between a Venus-like world and an Earth-like world, although recently published works on TRAPPIST-1 b and c raise concerns that TRAPPIST-1 d may be similarly devoid of a substantial atmosphere. TRAPPIST-1 d is also relatively understudied in comparison with TRAPPIST-1 e. The latter has generally appeared to be within the habitable zone of most atmospheric modeling studies. Assuming that TRAPPIST-1 d still retains a substantial atmosphere, we demonstrate via a series of 3D general circulation model experiments using a dynamic ocean that the planet could reside within the habitable zone in a narrow parameter space. At the same time, it could also be an exo-Venus or exo-Dead-type world or in transition between between one of these. Studies like this can help distinguish between these types of worlds.

\end{abstract}


\section{Introduction}\label{sec:intro}

The TRAPPIST-1 system \citep[e.g.][]{Gillon2017,Turbet2020,Ducrot2020,Bolmont2023,Agol2021} is one of the most exciting nearby planetary systems with seven transiting terrestrial planets. It promises to test our theories regarding terrestrial atmospheric evolution in M-dwarf systems, and it has already delivered surprising JWST observations of the two innermost planets \citep{Greene2023,Zieba2023}. TRAPPIST-1 b and c appear to lack substantial atmospheres, although the results for TRAPPIST-1 c remain under debate \citep[e.g.][]{Lincowski2023}. Other systems appear to produce similar results for close-in planets, e.g. GJ 367b \citep{Zhang2024} and GL 486b \cite{Mansfield2024}.

One of the key ingredients for traditional terrestrial planet habitability studies is whether the planet possesses surface water reservoirs. The TRAPPIST-1 system rocky planets may be amenable to substantial water delivery \citep[e.g.][]{Raymond2006,Raymond2007}. In addition, there is no evidence of any large outer planets that may block water delivery \citep[e.g.][]{Sabotta2021,Teske2022}, although other studies show that outer planets may enhance the impact rates of asteroids for solar system terrestrial planets \citep[e.g.][]{Horner2008}. The water mass fraction of TRAPPIST-1 d remains difficult to constrain \citep{Agol2021,Unterborn2018,Quarles2017} alongside the amount of water that may have condensed out on the surface assuming a temperate climate.

There are a number of studies that raise concerns about the ability of planets to retain surface water in TRAPPIST-1-type systems orbiting ultracool dwarf stars, where the magma ocean \citep[e.g.][]{Moore2023} and atmospheric escape (due to a number of processes) play important roles \citep[e.g.][]{Lammer2011,Bolmont2017,Airapetian2020,Fleming2020,Krissansen2023,Gialluca2024}.  Yet in order to be  ``habitable" a planet needs to condense water out on the surface at the end of the magma ocean phase. For TRAPPIST-1 d some models may not be favorable \citep[e.g.][]{Hamano2015,Turbet2023} alongside that of Venus or Venus-like worlds \citep{Turbet2021}.

TRAPPIST-1 d has an orbital period of $\sim$4.05 Earth days \citep{Gillon2017,Agol2021}. Assuming that it is tidally locked, this puts the planet in a fast rotator regime \citep[e.g.][]{Yang2014,Kopparapu2016,Way2018}. For TRAPPIST-1 d its Rossby number (Rossby radius of deformation divided by the planet's radius) is $\sim$1 putting it in the fast rotating regime, whereas values for simulations like those of \cite{Yang2019} are much greater than 1 putting them in the slow regime (day length $>$ 32 Earth days). For slow rotators (e.g., $<$ 32 Earth days; \citealt{Yang2014,Way2018}) the substellar cloud feedback may keep the planet cool even at quite high insolations, but that feedback is muted in TRAPPIST-1 d. Regardless, TRAPPIST-1 d's insolation S0X$\sim$1.115--1.143 S$_{\earth}$ \citep{Gillon2017,Turbet2020,Agol2021} places it near the boundary of the inner edge of the habitable zone in 1D studies \citep[e.g.][]{Kopparapu2013,Kopparapu2014,Kopparapu2016,Lincowski2018}.

Part of the lack of detailed studies of TRAPPIST-1 d's hypothetical climate (in comparison with that of TRAPPIST-1 e) may be a misconception about its mass and the higher insolation originally published in \cite{Gillon2017}. While it is less massive than Earth (0.3--0.41 M$_{\earth}$; \citealt{Gillon2017,Turbet2020,Agol2021}), it is 3--4 times the mass of Mars (0.107 M$_{\earth}$)\footnote{https://nssdc.gsfc.nasa.gov/planetary/factsheet/marsfact.html}. Its mass places it just inside the collisional atmosphere side of the cosmic shoreline \citep{Zahnle2017}. While TRAPPIST-1 e studies place it within the habitable zone \citep[e.g.][]{Wolf2017,Turbet2018,Lincowski2018,Lin2021,Krissansen2023} recent 3D GCM studies find that TRAPPIST-1 d is likely an exo-Venus \citep{Wolf2017,Turbet2018}, although \cite{Turbet2018} do state that ``if TRAPPIST 1d is able somehow to sustain a thick, highly reflective water cloud cover near the substellar region, it could sustain surface liquid water global oceans." However, as mentioned above, TRAPPIST-1 d is in the fast rotator regime of atmospheric dynamics; hence, such a substellar cloud deck needs to be tested.
For example, \cite{Kopparapu2017} Table 1 for a 2600K stellar type with an orbital period of 4.07 (very close to TRAPPIST-1 d's 4.05) they state that their model top H$_{2}$O mixing ratio (1 mb) is ``runaway, no water-loss limit." This is also demonstrated in Figures 2, 3 and 4 from the same work. This is in contrast to the earlier work by \cite{Yang2013} who appear to have used the incorrect orbital period (60 days) in their baseline M-star (3400 K) simulations and used an older version of the NCAR model than \cite{Kopparapu2017} (the former using CAM3 and the latter CAM4; see Figure 2 in \citealt{Kopparapu2017}). A 60-day orbital period will place this particular planet in the slow rotator regime. Finally, in a 3D GCM study, \cite{Kopparapu2017} state that TRAPPIST-1 d ``is likely in a runaway state" based on their analysis.

A number of studies have examined effects such as tidal dissipation and resultant surface heating fluxes \citep[e.g.][]{Barr2018,Makarov2018,Dobos2019,Bansal2023}. 
\cite{Barr2018} estimate modest tidal heating fluxes for TRAPPIST-1 d of $\sim$ 20$\times$ higher than modern Earth's, or $\sim$0.16 W m$^{-2}$. Such a modest amount of surface heating would not have a large net atmospheric heating effect. 
Subsequent work by \cite{Dobos2019} almost doubles the surface flux estimates of \cite{Barr2018} with a value of $\sim$0.26 W m$^{-2}$, but again this is a minor fraction of the received insolation of 1556 W m$^{-2}$.

Authors have examined the effects that stellar flares and coronal mass ejections (CMEs) may have on atmospheric erosion in low-mass M-star systems like TRAPPIST-1  \citep[e.g.][]{Khodachenko2007,Airapetian2020,DoAmaral2022} in unmagnetized \citep[e.g.][]{Tilley2019,France2020} and also magnetized planets \citep[e.g.][]{Dong2019}. Given our lack of knowledge as to whether any TRAPPIST-1 planet has an intrinsic magnetic field, such studies cover unconstrained parameter space and are extremely valuable considering that the TRAPPIST-1 star is highly active \citep{Howard2023}. Other authors have pointed out the effects that flares associated with CMEs might have on the internal dynamics of planets in the TRAPPIST-1 system. \cite{Grayver2022} have shown that internal heating rates generated from such interactions may be high enough to increase volcanism and possibly counteract atmospheric escape with associated  outgassing. In addition, mere movement of an object through a stellar magnetic field may induce internal induction heating also capable of producing increased volcanism \citep[e.g.][]{Kislyakova2017,Bromley2019}. \cite{Kislyakova2017} state that the outgassing of greenhouse gases may amount to as much as ``several tens of bar, mainly for Trappist 1b, 1c, 1d and 1e."

In general, planets orbiting M-dwarf stars may experience atmospheric loss from high stellar winds and high X-ray and ultraviolet fluxes \citep[e.g.][]{Lammer2003,Cohen2014,Cohen2015,GarciaSage2017,Garraffo2017}. Recent work by \cite{VanLooveren2024} has shown that all of the TRAPPIST-1 planets will have escape rates resulting in the loss of an entire Earth ocean in less than 1 Gyr. These estimates do not include loss from CMEs (mentioned above), which can increase escape rates by orders of magnitude \citep[e.g.][Section 2.3]{Way2023} higher than the estimates given in \cite{VanLooveren2024}. In the \cite{VanLooveren2024} scenario, given that the TRAPPIST-1 star is $\sim$ 8 Gyr old, this means a minimum of eight Earth oceans could have been lost. This makes the possibility of an atmosphere today challenging, assuming similar initial volatile inventories compared to that of Earth. In addition, there remains contentious debate about whether a magnetic field enhances or inhibits atmospheric escape \citep[e.g.][]{GarciaSage2017,Gunell2018,Gronoff2020,Gillmann2022,Way2023}. 

To balance atmospheric loss, estimates of the inventory and outgassing rates of volatiles are required \citep[e.g.][]{Unterborn2018,Unterborn2022}. Provided that sufficient volatiles exist and the interior retains enough energy \citep[e.g.][]{FoleySmye2018} to support a planetary-wide carbon cycle, TRAPPIST-1 d and e may be able to maintain a temperate climate over long timescales \citep[e.g.][]{Unterborn2022}. One might assume that a modern Earth-like carbonate-silicate cycle \citep[e.g.][]{Walker1981,Sleep2001,Kasting2003,Stewart2019,Graham2020,Lehmer2020} is required. But one needs to discriminate \emph{modern day} plate tectonics and its associated carbonate-silicate cycle that relies on 2/5 of Earth having subaerial crust with its implications for habitability \citep[e.g.][]{Rushby2018,Honing2023}. This contrasts with the fact that Earth has had some form of volatile cycling through most of the past 4+ Gyr with varying amounts of exposed land \citep[e.g.][]{Kasting2003,Cawood2022} while our Sun's luminosity has changed $\sim$25\% \citep[e.g.][]{Gough1981,Claire2012}. On top of changes in how volatile cycling has operated over the eons alongside CO$_2$ partial pressures, other studies have attempted to constrain surface pressures in the Archean, finding values that range from 0.23 to 2 bar \citep[][]{Goldblatt2009,Som2012,Marty2013,Som2016}. Exploring non-1-bar atmospheric surface pressures in 3D models in the context of the habitable zone remains an under explored area.

Given the uncertainty around TRAPPIST-1 d and the relative lack of detailed 3D GCM studies, we thought it worthwhile to add to the discussion with a new suite of 3D GCM experiments to help determine whether TRAPPIST-1 d resides in the canonical habitable zone \cite[e.g.][]{Kasting1993,Kopparapu2016}, resides in the Venus Zone \citep[e.g.][]{Kane2014}, or is a world devoid of a detectable atmosphere (exo-Dead) as may be the case for its cousins TRAPPIST-1 b and c \citep{Greene2023,Zieba2023}.

\section{Methods}\label{sec:methods}

We utilize the Planet 1.0 version of the ROCKE-3D (R3D) GCM in this work (referred to as R3DP1). Technical details about the R3DP1 model used herein can be found in \cite{Way2017}. In summary, R3DP1 is a 3D general purpose planetary GCM that couples atmosphere, surface, and ocean models into one. We run R3DP1 at a latitude $\times$ longitude resolution of 4 $\times$ 5$\degr$, with 40 atmospheric layers, 6 ground layers, and up to 13 ocean layers. The ocean used in this model is a dynamic ocean with depths, depending on the configuration, up to 310 m (Sims 10, 11), 899 m (Sims 04--09, 16, 17, 21--23), and 1940 m (Sims 03, 15, 20). The dynamic ocean plays an important and undervalued role in both oceanic and atmospheric heat redistribution. This was shown previously in \cite{DelGenio2019} for Proxima Centauri b, where the global mean surface temperature of a dynamic ocean simulation was 16$\degr$ warmer than the equivalent Q-flux/slab ocean version (see their Figures 1(a) and 1(b)). Unlike what are termed Q-flux, thermodynamic, or slab oceans \citep[][see their Section 2.2.2 for a detailed description]{Way2017}, a dynamic ocean will not display an artifact like a so-called eyeball world \citep{Pierrehumbert2010} in an aquaplanet setup, and the ``temperate terminator" (if it exists) will be more properly quantified. In this work we model TRAPPIST-1 d with two different atmospheric surface pressures (500 mb and 1000 mb). The choice of lower atmospheric pressure experiments is driven by work showing that such pressures may have occurred during the Archean period in Earth's history \citep{Som2012,Marty2013,Som2016}.
We also model TRAPPIST-1 e in a few 1000 mb cases for comparison purposes. Following the \citet[hereafter W2017]{Wolf2017} and \cite{Turbet2018} papers cited above, we mostly focus on atmospheres dominated by N$_2$, CO$_2$ and H$_2$O, consistent with atmospheres in traditional habitable zone studies \citep[e.g.][]{Kasting1993,Kopparapu2013}. In most cases we model a N$_2$-dominated atmosphere with 400 ppmv CO$_2$. For the sake of comparison we include a preindustrial (1850) Earth simulation 
%that includes aerosols, O$_2$, O$_3$ and CO$_2$=285.2ppmv N$_2$O=0.2754 and CH$_4$=0.791
(Sim 12 in Table 1). In two of the TRAPPIST-1 d 1000mb cases we model N$_2$-dominated atmospheres following \cite{Wolf2017,Chaverot2023} with CO$_2$ set to zero and 1\% (denoted as W1 and W2 in Table 1). Topography, land/sea mask, and surface water inventories follow the work of \citet[][see their Section 6 for detailed descriptions]{Way2020} where four were chosen for this study (A--D; see below), ranging from very little surface water (20 cm in the soil layers = Arid-Venus) to a planet covered in water (aquaplanet or AP). We added two additional setups from \citet[][E and F]{DelGenio2019} that use a modern-Earth-like land/sea mask with the substellar point placed over Africa (Day-Land) and the other placed over the Pacific Ocean (Day-Ocean). This provides a contrast in surface albedo, where Day-Land has a higher surface albedo at the substellar point than Day-Ocean. It is also an interesting contrast since the Day-Ocean simulation is bordered on the west by the Americas and on the east by Austral-Asia providing a western and an eastern boundary for ocean heat transport. This boundary affect should be more muted in the Day-Land case.

\begin{enumerate}[A)]
    \item Arid-Venus. Modern Venus topography with 20 cm of water emplaced within the soil layers of the model.
    \item 10 m-Venus. Modern Venus topography with 10 m global equivalent layer (GEL) deposited as lakes in the lowest topographic regions.
    \item 310 m-Venus. Modern Venus topography with 310m GEL deposited in ocean basins and lakes in the lowest topographic regions.
    \item Aquaplanet. A 900 m deep ocean emplaced globally except at the south pole point where a land grid cell of zero height remains.
    \item Day-Ocean. Modern-Earth-like setup with the substellar point located over the Pacific Ocean.
    \item Day-Land. Modern-Earth-like setup with the substellar point located over the continent of Africa.
\end{enumerate}

At model start most land albedos are set to 0.2 while the soil is a 50/50 mix of clay and sand as used in similar studies \citep{Yang2014,Way2018}. The exception to this rule are the Arid-Venus cases, where we use 100\% sand with an albedo of 0.3. There is no land ice at model start, but if any accumulates, its initial broadband albedo (visible+IR) will be 0.8 (0.95 in the visible) and below depending on the age.  The same applies to any sea ice. The ocean albedo is 0.15 at model start. However, all albedos have a spectral dependence that is discussed in detail in \cite{Way2017}.

With the exception of run W1,I in Table 1 (see below) we use the latest estimated insolation for TRAPPIST-1 d from \cite{Agol2021} S0X=1.115 S$_{\earth}$ $\sim$ 1517.5 W m$^{-2}$. This value was used in the recently published 1D modeling work of \cite{Meadows2023}, which is $\sim$1\% lower that that used in the 3D GCM studies of \cite{Wolf2017,Turbet2018,Rushby2020} of S0X=1.143 S$_{\earth}$ $\sim$ 1561 W m$^{-2}$. We use a sidereal orbital period of 4.05 Earth days from \cite{Agol2021,Gillon2017}. We used radius R=0.77 R$_{\earth}$, and mass M=0.41 M$_{\earth}$ giving a surface gravity of g = 6.75 m s$^{-2}$ (0.688 g$_{\earth}$) from \cite{Gillon2017}. 
For TRAPPIST-1 e we also used the estimates from \cite{Agol2021} for insolation S0X=0.646 S$_{\earth}$ $\sim$ 879 W m$^{-2}$ and an orbital period of 6.1 Earth days. Radius R=0.92 R$_{\earth}$ and mass M=0.62 M$_{\earth}$ give a surface gravity of g = 8.01 m s$^{-2}$ (0.817  g$_{\earth}$) from \cite{Gillon2017}. 

We utilize the 12/21\footnote{See ``More robust spectral files" https://simplex.giss.nasa.gov/gcm/ROCKE-3D/Spectralfiles.html} SOCRATES \citep{Edwards_Slingo1996,Manners2012,Amundsen2016} radiative transfer tables, which have 12 bands in the longwave and 21 bands in the shortwave. These files use the MT\_CKD 3.0 water vapor continuum \citep{Mlawer2012}. These are the same tables and TRAPPIST-1 stellar spectrum that were used for the Ben1 ROCKE-3D simulation in \cite{Turbet2022}, a 2600K BT-Settl model with Fe/H=0 \citep{Allard2012}. This is the same spectrum used in the work of W2017 discussed below.

We do two experiments (05 and 17) with a 3:2 spin--orbit rotation rate since work indicates that states that are not tidally locked may be possible \citep[e.g.][]{Correia2014,Makarov2018,Colose2021,Valente2022,Revol2023,Revol2024}. Regardless, simulations with a 3:2 spin--orbit state provide a nice contrast to their 1:1 tidally locked aquaplanet counterparts (04 and 16). Obliquity is set to zero (normal to the orbital plane) along with eccentricity = 0 in all simulations to limit our parameter space and make comparisons between simulations easier. This is also the most probable scenario given the TRAPPIST-1 system's high coplanar configuration and high tidal damping \citep{Agol2021}.

% Ran DayLand configuration 3 different times plotted together in Trappist1d1barN2C400DayLand.jpg
% Given these results (nrad/tsurf) in comparision with DayOcean we have to use DayLnd4:
% 1.) Trappist1d_1bar_N2_C400_DayLand: w/o any fixes (nrad/tsurf=0.8/35)
% 2.) Trapp1d_1B_N2_C400_DayLnd3: w/CLOUDS_BUGFIX_2015 (nrad/tsurf=0.44/41)
% 3.) Trapp1d_1B_N2_C400_DayLnd4: w/CLOUDS_BUGFIX_2015 + CLOUDS_LIMIT_ETADN (nrad/tsurf=–0.08/44.00)

% Ran DayOcean configuration 3 different times plotted together in Trappist1d1barN2C400DayOcean.jpg
% Given these results (nrad/tsurf) in comparision with DayOcean we have to use DayOcn4:
% 1.) Trappist1d_1bar_N2_C400_DayOcean: w/o any fixes (nrad/tsurf=1.7/43)
% 2.) Trapp1d_1B_N2_C400_DayOcn3: w/CLOUDS_BUGFIX_2015 (nrad/tsurf=2.4/42)
% 3.) Trapp1d_1B_N2_C400_DayOcn4: w/CLOUDS_BUGFIX_2015 + CLOUDS_LIMIT_ETADN (nrad/tsurf=0.11/46)

% Ran Aquaplanet tidally locked in several configurations and plotted them in Trappist1d1barN2C400AP.jpg
% Given these results and the fixes we have to use AP4:
% 1.) ANN4501-5000.aijTrappist1d_1bar_N2_C400_AP.nc (nrad/tsurf=5.7/35)
% 2.) ANN9001-9990.aijTrapp1d_1B_N2_C400_AP3.nc w/CLOUDS_BUGFIX_2015 (nrad/tsurf=5.1/36)
% 3.) ANN34001-34990.aijTrapp1d_1B_N2_C400_AP4.nc w/CLOUDS_BUGFIX_2015 + CLOUDS_LIMIT_ETADN (nrad/tsurf=3.3/39)
% The latter is still climbing after 35000 orbits, but we end it here and highlight it red.

% Table 01
\begin{deluxetable}{|l|l|r|r|r|r|r|r|r|r|r|r|r|r|r|r}
\tabletypesize{\scriptsize}
\tablecaption{List of Experimental Results\label{table:exp}}
\tablehead{
\multicolumn{1}{|l|}{Sim} & \multicolumn{1}{|c|}{BC\tablenotemark{\tiny a}} & \multicolumn{1}{|c|}{Orbits\tablenotemark{\tiny b}}   & \multicolumn{1}{|c|}{P\tablenotemark{\tiny c}} & \multicolumn{1}{|c|}{Bal\tablenotemark{\tiny d}} & \multicolumn{1}{|c|}{T$_{surf}$\tablenotemark{\tiny e}} & \multicolumn{1}{|c|}{Alb\tablenotemark{\tiny f}} & \multicolumn{1}{|c|}{QT\tablenotemark{\tiny g}} & \multicolumn{1}{|c|}{W\tablenotemark{\tiny h}} & \multicolumn{1}{|c|}{CW\tablenotemark{\tiny i}} & \multicolumn{1}{|c|}{SCRF\tablenotemark{\tiny j}} & \multicolumn{1}{|c|}{LCRF\tablenotemark{\tiny k}} & \multicolumn{1}{|c|}{CRF\tablenotemark{\tiny l}} & \multicolumn{1}{|c|}{CL\tablenotemark{\tiny m}} & \multicolumn{1}{|c|}{Hab\tablenotemark{\tiny n}} \\
%\multicolumn{1}{|l|}{} & \multicolumn{1}{|c|}{} & \multicolumn{1}{|c|}{}& \multicolumn{1}{|c|}{(mb)} & \multicolumn{1}{|c|}{W m$^{-2}$} & \multicolumn{1}{|c|}{(K)} & \multicolumn{1}{|c|}{(\%)} & \multicolumn{1}{|c|}{v v$^{-1}$} & \multicolumn{1}{|c|}{kg m$^{-2}$} & \multicolumn{1}{|c|}{kg m$^{-2}$} & \multicolumn{1}{|c|}{W m$^{-2}$} & \multicolumn{1}{|c|}{W m$^{-2}$} & \multicolumn{1}{|c|}{W m$^{-2}$} & \multicolumn{1}{|c|}{\%} & \multicolumn{1}{|c|}{\%}\\
\hline
\multicolumn{15}{|c|}{TRAPPIST-1 d}
}
\startdata
01 & Arid-Venus & 34999 & 1000 &  0.02 & 286(229/364)&  21 & 2e-5 &   3 &  2e-3 & -2.03 &  6.12 &  4( -6/ 73)& 24& 48\\ % 01_ANN34000-34999.aijTrappist1d_1bar_N2_C400_Arid.nc
02 & 10m-Venus & 34990 & 1000 & -0.00 & 307(265/379)&  15 & 3e-5 &  39 &  0.08 & -21.81 & 22.45 &  1(-90/ 89)& 29& 96\\ % 02_ANN34001-34990.aijTrappist1d_1bar_N2_C400_10m.nc
03 & 310m-Venus & 33960 & 1000 &  1.70 & 321(290/365)&  25 & 9e-5 & 447 &  0.82 & -83.76 & 31.95 &-52(-378/ 93)& 67&100\\ % 03_ANN33001-33960.aijTrappist1d_1bar_N2_C400_310m.nc
\cellcolor{red!15}04 & AquaPlanet & 34990 & 1000 &  3.28 & 312(295/331)&  18 & 1e-4 & 347 &  0.76 & -63.26 & 22.23 &-41(-207/ 49)& 59&100\\ % 04_ANN34001-34990.aijTrapp1d_1B_N2_C400_AP4.nc
05 & AP(3:2) & 30000 & 1000 &  1.17 & 320(308/330)&  21 & 1e-5 & 485 &  0.73 & -74.73 & 24.57 &-50(-78/ 26)& 62&100\\ % 05_ANN29001-29990.aijTrapp1d_1B_N2_C400_AP32v5.nc
06 & AP(W1) & 30000 & 1000 & -0.13 & 296(270/319)&  23 & 9e-6 & 144 &  0.59 & -78.77 & 26.79 &-52(-270/ 56)& 74&100\\ % 06_ANN29001-30000.aijTrappist1d_1bar_N2_W1_AP.nc
07 & AP(W1,I) & 30000 & 1000 & -0.09 & 299(277/323)&  24 & 3e-5 & 185 &  0.63 & -83.94 & 27.27 &-57(-298/ 58)& 74&100\\ % 07_ANN29001-30000.aijTrappist1d_1bar_N2_W1_AP2.nc
08 & AP(W1,QF) & 4990 & 1000 & -0.00 & 288(248/326)&  25 & 2e-5 & 180 &  0.64 & -83.85 & 24.80 &-59(-283/ 37)& 75& 73\\ % 08_ANN4001-4990.aijTrapp1d_1B_N2_W1_AP_QF.nc
\cellcolor{red!15}09 & AP(W2) & 7000 & 1000 &  8.52 & 311(299/322)&  28 & 1e-3 & 176 &  0.39 & -92.69 & 26.74 &-66(-372/ 37)& 78&100\\ % 09_ANN6001-7000.aijTrappist1d_1bar_N2_W2_AP2.nc
10 & Day-Ocean & 25000 & 1000 &  0.11 & 319(289/348)&  25 & 1e-3 & 678 &  0.86 & -90.07 & 32.65 &-57(-323/ 86)& 72&100\\ % 10_ANN24001-25000.aijTrapp1d_1B_N2_C400_DayOcn4.nc
11 & Day-Land & 25000 & 1000 &  1.38 & 317(290/366)&  26 & 1e-3 & 441 &  0.76 & -89.61 & 34.33 &-55(-366/ 86)& 71&100\\ % 11_ANN24001-25000.aijTrapp1d_1B_N2_C400_DayLnd4.nc
12 & Earth & 8999 & 984 &  0.02 & 286(205/308)&  30 & 4e-6 &  21 &  0.45 & -50.36 & 20.85 &-30(-89/ 16)& 55& 83\\ % 12_ANN8500-8999.aijE1oM20_Test2.nc
\hline
13 & Arid-Venus & 34990 & 500 &  0.03 & 259(185/368)&  25 & 3e-5 &   0.3 &  2e-4 & -0.52 &  5.92 &  5( -2/156)& 27& 38\\ % 13_ANN34001-34990.aijTrappist1d_500mb_Arid_Venus.nc
14 & 10m-Venus & 34990 & 500 & -0.02 & 292(235/381)&  14 & 7e-4 &   9 &  0.01 & -5.33 & 14.14 &  9(-26/119)& 33& 47\\ % 14_ANN34001-34990.aijTrappist1d_500mb_10m_Venus.nc
15 & 310m-Venus & 10000 & 500 &  1.06 & 294(239/346)&  33 & 1e-2 & 190 &  0.45 & -108.81 & 38.04 &-71(-399/107)& 87& 83\\ % 15_ANN9001-10000.aijTrappist1d_500mb_310m_Venus2.nc
16 & AquaPlanet & 8990 & 500 & -2.56 & 295(268/318)&  32 & 7e-3 & 192 &  0.43 & -112.40 & 30.62 &-82(-382/ 34)& 89& 94\\ % 16_ANN8001-8990.aijTrappist1d_500mb_AP_TL.nc
17 & AP(3:2) & 8990 & 500 &  2.50 & 300(274/311)&  29 & 5e-4 & 162 &  0.42 & -102.65 & 27.43 &-75(-129/ 25)& 83&100\\ % 17_ANN8001-8990.aijTrappist1d_500mb_AP_32.nc
\hline
\multicolumn{15}{|c|}{TRAPPIST-1 e}\\
\hline
18 & Arid-Venus & 5000 & 1000 &  0.03 & 233(182/302)&  26 & 6e-7 & 0.02 &  0.01 & -2.23 &  3.58 &  1(-48/ 40)& 54& 19\\ % 18_ANN4500-4999.aijTrappist1e_01.nc
19 & 10m-Venus & 5000 & 1000 & -0.33 & 244(197/306)&  19 & 7e-7 &   1 &  0.03 & -10.20 &  7.63 & -3(-185/ 29)& 71& 25\\ % 19_ANN4500-4999.aijTrappist1e_02.nc
20 & 310m-Venus & 3958 & 1000 & -1.44 & 244(202/293)&  23 & 4e-7 &   3 &  0.08 & -25.92 & 10.19 &-16(-210/ 54)& 83& 19\\ % 20_ANN3458-3958.aijTrappist1e_03.nc
21 & AquaPlanet & 5000 & 1000 & -0.32 & 252(213/276)&  19 & 5e-7 &   3 &  0.04 & -22.04 & 10.48 &-12(-190/ 19)& 88& 14\\ % 21_ANN4500-4999.aijTrappist1e_04.nc
\cellcolor{red!15}22 & AP(W2) & 10989 & 1000 & -0.57 & 258(224/283)&  19 & 1e-6 &   4 &  0.06 & -23.34 &  9.32 &-14(-170/ 18)& 83& 23\\ % 22_ANN10000-10989.aijTrappist1e_05.nc
23 & AP(CO2) & 3000 & 1000 &  1.43 & 292(285/299)&  18 & 5e-5 &  27 &  0.18 & -34.48 &  8.23 &-26(-128/  8)& 76&100\\ % 23_ANN2500-2999.aijTrappist1e_1bar_CO2_AP.nc
\hline
\enddata
%\end{tabular}
\tiny
Runs 04, 09 and 22 are not believed to be in net radiative balance (NRB) at the top of the atmosphere (TOA) either because of trends in the surface temperature (see the Appendix) or as a result of premature model termination due to other factors (e.g., Sim 20's ocean froze to the bottom in orbit number 10,990).
\tablenotetext{a}{Boundary conditions: N$_2$-dominated atmospheres with 400 ppmv CO$_2$ except for W1 = after W2017 (CO$_{2}$ = 0ppmv), W1,QF = W1 with Qflux = 0/slab ocean, W1,I = W1 with \cite{Gillon2017} higher insolation, and W2 = after W2017 (CO$_{2}$ = 1\%). QF: Qflux = 0 slab ocean used. (3:2) indicated the 3:2 spin--orbit rotation rate, and all others are tidally locked except for the Earth simulation (12), which uses modern-day Earth's rotation rate. See text for additional information.}
\tablenotetext{b}{Number of orbits completed.}
\tablenotetext{c}{Surface pressure (mbar).}
\tablenotetext{d}{NRB at the TOA (W m$^{-2}$). Same averaging as for surface temperature.}
\tablenotetext{e}{Global mean surface temperature (min/max) averaged over 1000 orbits (K). If the orbits are $\leq$7000 then we average over 500 orbits.}
\tablenotetext{f}{Planetary albedo (\%)}
\tablenotetext{g}{Specific humidity (Q) in top (T) pressure layer (v v$^{-1}$).}
\tablenotetext{h}{Vertical-integrated water vapor content (kg m$^{-2}$).}
\tablenotetext{i}{Vertical-integrated cloud water vapor content (kg m$^{-2}$).}
\tablenotetext{j}{Shortwave cloud radiative forcing (W m$^{-2}$).}
\tablenotetext{k}{Longwave cloud radiative forcing (W m$^{-2}$).}
\tablenotetext{l}{Net (shortwave + longwave) cloud radiative forcing (min/max) (W m$^{-2}$).}
\tablenotetext{m}{Total cloud coverage (\%).}
\tablenotetext{n}{Surface habitability percent \citep{Spiegel2008}.}
\end{deluxetable}

\begin{figure}[ht!]
\centering
\includegraphics[scale=0.2]{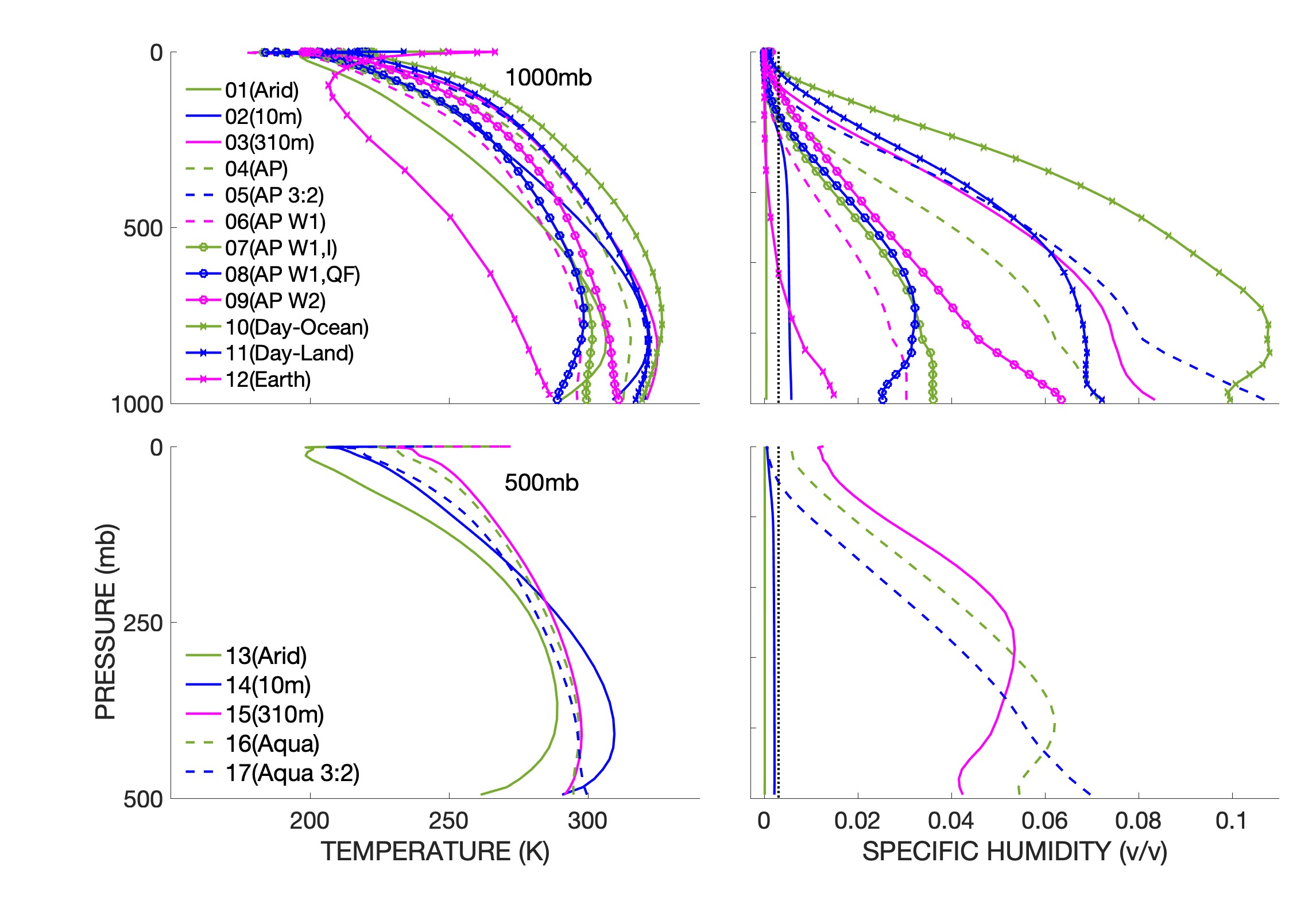}
\caption{TRAPPIST-1 d temperature and specific humidity  as a function of model pressure layers. Top row: 1000 mb; bottom row: 500mb. The vertical dotted black line in the right panels represents the moist greenhouse limit of 3x10$^{-3}$ (v/v) from \cite{Kasting1993}. We include a modern Earth simulation as a comparison, but note that it uses 20 vertical layers, whereas all other simulations use 40.}
\label{fig:1}
\end{figure}

\begin{figure}[ht!]
\centering
\includegraphics[scale=.55]{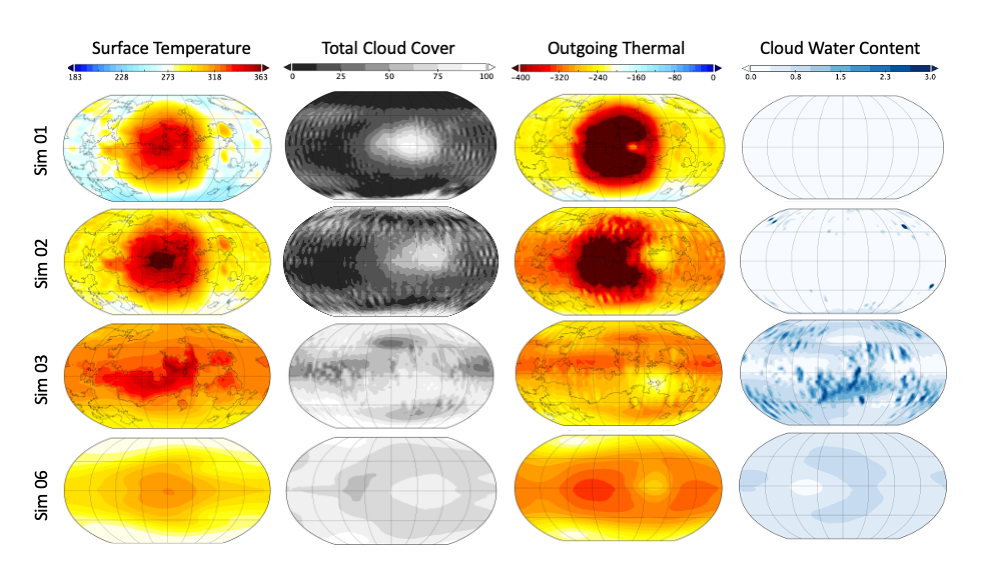}
\caption{Surface temperature (straddling 0$\degr$C), total cloud cover (\%), net outgoing thermal radiation (W m$^{-2}$), and cloud water content (kg m$^{-2}$) for four of the simulations listed in Table \ref{table:exp}. Simulations are displayed from dry (Sim 01: Arid-Venus), semidry 10 m GEL (Sim 02: 10 m-Venus), Earth-like 310 m GEL (Sim 03: 310 m-Venus) and an aquaplanet (Sim 06). Sims 01--03 are 1 bar N$_2$ dominated with 400ppmv CO$_2$. Sim 06 is N$_2$-dominated, but the only greenhouse gas is H$_2$O following \cite{Wolf2017}. Cloud coverage is fairly low in Sims 01--02 as expected given limited surface reservoirs and cloud water content (rightmost column), and is much higher for Sims 03 and 06.}
\label{fig:2}
\end{figure}

The column labeled Q$_{top}$ in Table \ref{table:exp} estimates whether the simulation has approached the \cite{Kasting1993} moist greenhouse limit of f(H$_2$O)=3x10$^{-3}$ (v v$^{-1}$) in the upper atmosphere. We extract values from the highest atmospheric layer of the model following that of Figure 2 in \cite{Wolf2015}.  This is necessary since it is not enough to simply take the value above the tropopause in the stratosphere as done in 1D radiative convective models, where stratospheric temperatures and water vapor are often fixed constant values from the tropopause to the top of the model.

\section{Discussion}\label{sec:discussion}

Given the uncertainty expressed in previous work about the nature of TRAPPIST-1 d's climate \cite[e.g][]{Turbet2018}, it is instructive to compare our results with some of those in the existing literature.

A number of excellent 1D modeling studies have looked at TRAPPIST-1 d 
\citep[e.g.][]{Lincowski2018,Meadows2023} that we do not compare in detail in this short Letter. Rather, we focus our discussion on comparison with 3D GCM studies. For example, \cite{Fujii2017} examined near-infrared-driven moist upper atmospheres for GJ 876 (M4V stellar T$_{eff}$ = 3129 K, whereas Trappist 1 T$_{eff}$ = 2559K). In one case they examined a world with an incident flux of S0X$=$1.2 times that of modern Earth and an orbital period of P $=$ 22 days (recall TRAPPIST-1 d  S0X $=$ 1.115 and P $=$ 4.05 days). They use an aquaplanet setup but stick to an Earth radius and surface gravity. Yet we only have one simulation (Sim 03) that approaches the 320 K (47$\degr$C) surface temperature necessary for the upper atmosphere specific humidity structure they describe in their Fig 2 for the GJ876 S$_{x}$ = 1.2 case. Our Sim 03 does not exhibit the same structure, and we remain below the \cite{Kasting1993} 3x10$^{-3}$ v v$^{-1}$ moist greenhouse limit (see Table \ref{table:exp} and Figure \ref{fig:1}). The extra insolation and different rotation rate and radius, combined with a hotter stellar effective temperature in their GJ876 S$_{x}$ = 1.2 case probably play the main roles in keeping our simulations distinct. 

\cite{Kopparapu2016} simulate an aquaplanet with a similar insolation (1600 W m$^{-2}$, vs. 1517.5 W m$^{-2}$ for TRAPPIST-1 d) and an orbital period of 9 days (vs. TRAPPIST-1 d's 4 days). 
Their simulation uses Earth's radius; hence, given the differing rotation periods and radii, their Rossby number (Rossby radius/planetary radius) will be different from those of TRAPPIST-1 d as discussed above. They also use a stellar T\_eff = 3300 K, while for TRAPPIST-1 T\_eff = 2500 K. They use a 1-bar N$_2$-dominated atmosphere with 1 ppmv CO$_2$. Our Sim 03 is the closest simulation, but we only have H$_2$O as our greenhouse gas (we are missing their 1 ppmv CO$_2$). Setting aside those differences, one might compare our Sim 06 with their simulation shown in \citet[][Figure 3, right column]{Kopparapu2016} . Their mean surface temperature is $\sim$256 K while that for  Sim 06 is 296 K (a difference of $\sim 40\degr$). The cloud cover patterns and cloud water column are not dissimilar, but clearly the differences in the model setups and cloud properties account for the 40$\degr$ difference. Below we discuss model cloud properties that may go some way to explaining this difference.

A more recent work that one might expect to be comparable to TRAPPIST-1 d is that of \cite{Chaverot2023}, who examine the runaway greenhouse transition. Their simulation labeled W1 is a similar setup to our Sim 06, but they use a solar spectrum, and as shown in \cite{Fujii2017}, there is a stark difference when using an M-dwarf versus a G-dwarf spectrum. None of our simulations look similar to the results in \cite{Chaverot2023}. If we stretch the comparison with \cite{Chaverot2023} in their Fig. 11, our cloud radiative forcing is generally to the left of their dotted line for 1 bar of water vapor, and our simulations would be in what they term ``stable states." The same applies to Figure 12 in \cite{Chaverot2023}, where again our albedos broadly mimic the same `stable states' behavior below surface temperatures of 320 K. The conclusion from comparing with both of these works is that all of our simulations of TRAPPIST-1 d are far from the runaway greenhouse state, while a few may be approaching the moist greenhouse in the case where the simulations are not quite in net radiative balance (NRB; see Table \ref{table:exp} and caption).

\cite{Rushby2020} modeled TRAPPIST-1 d as an arid planet with the same 3D GCM used in W2017. They used soil types of different albedos (see their Table 2). In our most comparable case called Arid-Venus (Sim 01) we use a 100\% sand soil composition but set the global land albedo to 0.3 in the visible and near-IR at model start. Our Arid-Venus setup is most comparable to the land planet granite soil albedo used in the work of \cite{Rushby2020}, which has VIS/IR values of 0.294/0.277. In this granite case the mean global, dayside and nightside temperatures were 244.5, 276.5, ad 210.6 K (--28.5$\degr$C, +3.5$\degr$C, --62.9$\degr$C) respectively, which are significantly lower than our values for Sim 01, with our global mean being 286 K (13$\degr$C; see Table \ref{table:exp}). Their top-of-the-Atmosphere (TOA) global mean albedo (what we term planetary albedo) is 0.256, while ours is 0.288 at model completion. One major difference may be due to the fact that we include 20 cm of water in all soil grid cells at model start. It does not appear that \cite{Rushby2020} do the same; hence, this water provides a modicum of additional H$_2$O greenhouse gas forcing to our atmosphere as seen in the formation of modest amounts of clouds in Figure \ref{fig:2} and Table \ref{table:exp} (column CL) with a small radiative forcing (column CRF).

Another possible 3D GCM comparison is Kepler 1649b which was modeled with R3DP1 \citep{Kane2018}.
This planet has an insolation of S0X=2.3 S$_{\earth}$, hence far higher than TRAPPIST-1 d. However, the authors attempted to find out where a planet in this system would enter the habitable zone, but the only simulation in NRB (Sim 10) has S0X = 1.4 S$_{\earth}$ and a 50-day orbital period, putting it in the slow rotator regime, and TRAPPIST-1 d is not in this dynamical regime. All simulations in this same work are in the moist greenhouse regime except Sim 8, which is not in NRB and has a modern day Earth insolation (lower than TRAPPIST-1 d).

Tidally locked arid worlds tend to have stark temperature contrasts between the permanent dayside and nightside, meaning that the mean surface temperature holds a complicated meaning (see Table 2 in \citealt{Rushby2020}). Still, our 1-bar Arid-Venus (Sim 01) dayside surface temperatures are nowhere near what one would characteristically define as Venus-like (364 K = 91$\degr$C at the substellar point, 279 K = $\sim$6$\degr$C near the terminator). On the other hand, the nightside has cooler temperatures, reaching values as low as 230 K = --43$\degr$C (still 20$\degr$ warmer than the \citealt{Rushby2020} granite simulation value of 210.1 K = --62.9$\degr$C). The day--night temperature contrast for the Arid-Venus land planet increases with decreasing surface pressure (Sim 01 vs. Sim 13), while the mean surface temperature increases with pressure, as one might expect. Yet even if arid/land planets at higher insolations have more temperate conditions than aquaplanets \citep[e.g.][]{Abe2011}, would an ``Arid-Venus" be capable of hosting life given the very limited water availability? In this sense an aridity index might be preferred \citep[e.g.][]{DelGenio2019ClimatesIII} over a simple temperature range \citep[e.g.][]{Spiegel2008} for assessing habitability. Given the brevity of this Letter, in Table 1 we calculate a simple surface habitable fraction \citep{Spiegel2008,Sparrman2020,Woodward2024} where habitable indicates the fraction of the surface with temperatures between 0 $<$ T $<$ 100$\degr$C. In general, TRAPPIST-1 d is more ``habitable" than Trappist 1e according to these results.

Sims 06, 07 and 08 are readily comparable to the previously published work of W2017 in particular their ``Planet d." Planet d is TRAPPIST-1 d with a 1-bar N$_2$-only atmosphere in an aquaplanet setup, and our Sims 06--08 are meant to compare to this simulation ``Planet d" is the only other published 3D GCM aquaplanet simulation of TRAPPIST-1 d. Simulations 06--08 are aquaplanets; hence, H$_2$O provides the only greenhouse gas. One major difference in outcomes between W2017 and our simulations is the mean global surface temperature (T$_{global}$).
In W2017 their value is T$_{global}$ = 382 K (109$\degr$C), whereas in our Sim 06 it is 296 K (23$\degr$C) a drastic difference where W2017 is +86$\degr$ warmer, but note that W2017's published ``Planet d" simulation is not in radiative balance (it is approaching a runaway greenhouse). Regardless, ``Planet d" uses a slightly higher insolation from the earlier published work of \cite{Gillon2017} (I = 1555.6 W m$^{-2}$) versus \cite{Agol2021} (I = 1517.5 W m$^{-2}$) in our simulations. To quantify the impact of the different insolations used we have used this higher insolation in Sim 07 which yields a T$_{global}$ = 299K (3$\degr$C warmer), reducing the difference from ``Planet d" to 83$\degr$C. But both Sim 06 and Sim 07 use a fully coupled ocean, and W2017 uses a slab ocean. To that end, we replicated Sim 06 with a slab ocean in Sim 08 producing T$_{global}$ = 288 K, but this makes the difference with ``Planet d" even larger (94$\degr$C).
These discrepancies may be due to some differences in  how the clouds are modeled. The Total Cloud Cover column for Sim 06 in Fig \ref{fig:2} shows clouds on the dayside that appear to be absent in W2017 Planet d (their Figure 4, bottom row). Although both simulations have some low-level nightside clouds. The dayside clouds (east of the substellar point) are high-level clouds with a correlating longwave (thermal) cloud radiative forcing (not shown). 

Another simulation that closely mimics our Sims 06--08 is found in the work of \citet[][hereafter W2019]{Wolf2019}. In their Figures 2--5 W2019 they model what they refer to as a ``Temperate" world orbiting a star similar to TRAPPIST-1 with a T$_{eff}$ = 2600 K. Their ``Temperate" planet\footnote{The simulation has the name t2600\_s1350\_p4.13214.cam.h0.avg\_n68.nc in their online archive 
\url{https://archive.org/download/SimulatedPhaseDependentSpectraOfTerrestrialAquaplanetsInMdwarfSystems} 
which includes the insolation and orbital period in the filename.}
is an Earth-sized aquaplanet (hence larger than TRAPPIST-1 d) with an orbital period of 4.13 days and an insolation of 1350 W m$^{-2}$ (S0X = 0.992) which is substantially lower than that used in our Sims 06--08 (1517.5--1561 W m$^{-2}$). They also use a slab ocean as in W2017. The mean surface temperature for their simulation is 300K, nearly the same as our Sim 07 (299 K). However, it is a bit warmer than our slab ocean version Sim 08 at 288 K (a 12$\degr$ difference). Regardless, even though the setup of the simulation is very distinct from Sims 06--08, it exhibits qualitatively similar temperature, cloud water vapor, and cloud cover.

The reason for the stark differences with W2017 and W2019 is not easy to ascertain, but let's discuss effects related to clouds that may account for some of the differences. It should be noted that cloud particle number in R3DP1 is fixed over the ocean surface for liquid water at $\sim$6x10$^{7}$ m$^{-3}$. 
For liquid water over land it is $\sim$1.7$\times$10$^{8}$ m$^{-3}$. For ice clouds it is the same regardless of land or ocean with a value of $\sim$6$\times$10$^{4}$ m$^{-3}$ \citep{Sergeev2022,DelGenio1996}.
Both Exocam (used in the W2017 and W2019 simulations) and ROCKE-3D cloud particle sizes are not specified; they are interactive, increasing as the amount of liquid water or ice in the cloud increases. There are a number of  cloud parameters whose tunings may have an influence. For example, there are two R3DP1 cloud parameters (WMUI(L)\_multipler)\footnote{Water Mass Unit Ice (Liquid).} that govern the threshold for when cloud ice (liquid water) converts to precipitation, which in turn drives how much cloud ice (liquid water) can exist in the atmosphere before it precipitates. This in turn impacts optical depth/radiation.
There are also R3DP1 radius multipliers on the ice and liquid water particle sizes. The WMUI(L) parameters are related to autoconversion of cloud to precipitation in general and can impact how much condensate can be suspended, which is far from trivial in terms of impacts on optical depth when these parameters are set to different values between different 3D GCMs (as they will be between Exocam and ROCKE-3D). For example, if the WMUs are set to high values and the radius multipliers are low (relative to some typical equivalent settings in some other model via their own ways of computing condensate and particle size), then that could jointly work to increase reflection of incoming radiation (planetary albedo). A proper intercomparison \citep[e.g.][]{Yang2019,Fauchez2020,Turbet2022} is probably necessary in order to resolve the discrepancies between Exocam and R3DP1 for simulations like these.

For TRAPPIST-1 e we can compare an aquaplanet setup with the W2017 simulation they refer to as ``Planet e Temperate" where they obtain T$_{global}$ = 290 K (17$\degr$C). This compares nicely to our Sim 21 with T$_{global}$ = 292 K (19.1$\degr$C). However, W2017 use a 1-bar N$_2$-dominated atmosphere with 0.4-bar CO$_2$, whereas our simulation is a 1-bar pure CO$_2$ atmosphere. It is also informative to compare our TRAPPIST-1 e simulations to R3DP1 simulations of Proxima Centauri b \citep{DelGenio2019}. Their simulation 10 has T$_{global}$ = 284K versus our Sim 21 at 292 K. Both simulations include a 1-bar pure CO$_2$ atmosphere. TRAPPIST-1 e and PCb are in similar environments in the sense that PCb has an insolation of 882 W m$^{-2}$ versus TRAPPIST-1 e's value of 879 W m$^{-2}$ while orbiting M-dwarfs with moderately different T$_{eff}$ = 2992 K (PC) versus 2566 K for TRAPPIST-1. 
The orbital periods of the planets differ, with the latter TRAPPIST-1 e at 6 days and PCb at 11 days, but the atmospheric dynamics place both in the fast rotator regime. For these reasons we believe that this comparison is not unreasonable. The difference in mean surface temperature of 8$\degr$ demonstrates some consistency in our modeling efforts for such worlds with insolations $\sim$65\% lower than modern Earth. Our TRAPPIST-1 e simulations are in line with other 3D GCM results \citep[e.g.][]{Wolf2017,Turbet2018}, but there is an understandable spread given the models different boundary conditions and slightly different insolations since we use \cite{Agol2021} S0X=0.646  S$_{\earth}$ and they use S0X=0.662 S$_{\earth}$ from \cite{Gillon2017}. As shown above, this $\sim$1\% difference amounted to a difference of a few degrees in mean surface temperature in the TRAPPIST-1 d case.

In general, the TRAPPIST-1 d 1-bar simulations are fairly temperate compared with previous work. As seen in Figure \ref{fig:2}, Sim 01 (Arid-Venus) has the fewest clouds of any simulation. This makes sense given the lack of surface water available to the atmosphere. But it also has the lowest surface temperature of all of the TRAPPIST-1 d 1-bar simulations. This is due to the lack of H$_2$O as a greenhouse gas, which is in line with previous work \citep[e.g.][]{Abe2011,Kodama2019}. We believe the reason for Sim 03 (310 m-Venus) being warmer than its aquaplanet counterparts (Sims 04 and 05) is likely due to the distribution of the continents, which restricts the dayside-nightside ocean heat transport. In an aquaplanet setup there are no barriers to this transport. But the ratio of the amount of land to ocean and the distribution of the continents also play an important role since Sims 10 and 11 are cooler than Sim 03 even though in Sim 10 (Day-Ocean) case the Americas serve as a boundary to the east and Austral-Asia to the west. Simulation 09 (W2) with 1\% CO$_2$ is difficult to compare with the others since it is far from radiative balance. In Table \ref{table:exp} we denote this balance in Column 5 (Bal). In R3DP1 it is desirable for Bal to be $\pm$0.2 W m$^{-2}$, but we consider Sims 03, 11, 16 and 17 to be almost in balance as their global mean temperatures appear to be stable over a 1000-orbit timescale. We highlight in red those simulations we do not consider in balance at values $\gtrsim$ $\pm$1.5 W m$^{-2}$.

In the case of the TRAPPIST-1 d 500 mb simulations the climates, as expected, are considerably cooler.  Sim 13 is even below freezing for the global mean surface temperature (Arid-Venus). We believe that Sims 16 and 17 are almost in radiative balance since their surface temperatures appear stable over thousands of orbits. We do not believe Sim 22 is in balance since its global mean surface temperature had not stabilized.

\section{Conclusion}
Is TRAPPIST-1 d an exo-Venus, an exo-Earth or an exo-Dead world? We have shown that, given the assumptions used herein, in many cases it resembles an exo-Earth type world, with mostly moderate temperatures similar to much of Earth's history. This conclusion should be tempered with the knowledge that the distinct climate outcomes for TRAPPIST-1 d between this study (temperate) and previous work (runaway) are likely attributed to cloud parameterization approaches in different 3D GCMs.  At the same time, in the Introduction we pointed out a number of reasons why it may actually be exo-Dead. As with all similar modeling studies, only definitive observational data can confirm or deny these scenarios. In follow-up work we will examine whether any of the scenarios modeled herein are discernible from each other with current space- or ground-based assets. TRAPPIST-1 d observations may confirm or refute a number of other theories, including atmospheric escape and tidal heating. For example, will the cosmic shoreline hypothesis \citep{Zahnle2017} hold up, and which models of tidal dissipation may be proved more right than wrong \citep[e.g.][]{Barr2018,Makarov2018,Dobos2019,Bansal2023}? Does TRAPPIST-1 d reside in the Venus zone \citep{Kane2014} and should it be included in recently published Venus zone demographics \citep{Ostberg2023}? We wait in excited anticipation for such observational confirmation.

\begin{acknowledgments}
The author thanks Eric Wolf, Gregory Elsaesser, Tony Del Genio, and Andrew Ackerman for general discussions and model development related to this work. Thanks to Guillaume Gronoff for discussions regarding atmospheric escape.
This work was supported by NASA's Nexus for Exoplanet System Science (NExSS).  Resources supporting this work were provided by the NASA High-End Computing Program through the NASA Center for Climate Simulation at Goddard Space Flight Center. M.J.W. acknowledges support from the GSFC Sellers Exoplanet Environments Collaboration (SEEC), which is funded by the NASA Planetary Science Division's Internal Scientist Funding Model, and ROCKE-3D which is funded by the NASA Planetary and Earth Science Divisions Internal Scientist Funding Model. The Model and associated data used to generate the figures and analysis herein can be downloaded from \url{https://doi.org/10.5281/zenodo.14740675} or from \url{https://doi.org/10.17605/OSF.IO/V743T}.
We would also like to thank the anonymous referee for their insightful feedback and comments that greatly improved the text.
\end{acknowledgments}
\newpage

\begin{appendix}\label{sec:appendix}

The NRB at the TOA in GCMs is a widely used metric as to whether the model as a whole is in equilibrium. In the case of a world with a deep fully coupled dynamic ocean, the time to equilibrium can take hundreds or thousands of model years. In the case of ROCKE-3D, we aim for this value to approach $\pm$ 0.2 W m$^{-2}$. However, a few of the models in this paper are far from this value (see Table \ref{table:exp}). To make it clear to the reader what the state of each model run is (beyond the values listed in Column (5) of Table \ref{table:exp}) we plot in Figure \ref{fig:nradtsurf}A the last 240 orbits of each simulation following Figure 6 in \cite{Kopparapu2017}. Sometimes the model wanders out of the $\pm$ 0.2 W m$^{-2}$ range, but the temperature may remain stable, and that is the case for most models in Figure \ref{fig:nradtsurf}B. In the THAI intercomparison project \citep[][Section 4]{Fauchez2020} they aimed for NRB to approach 1 W m$^{-2}$ when averaged over a time span sufficient to capture the interannual variability. For Earth models this is typically around 10 yr. In the THAI I intercomparison paper for the dry runs 10 orbits was deemed sufficient (61 Earth days) \citep[][Section 2]{Turbet2022}. We aim for 1000 orbits in the TRAPPIST-1 d case (900 orbits would be equivalent to $\sim$10 Earth years) given the deep fully coupled dynamic ocean used. However, since the TRAPPIST-1 d runs have zero obliquity and eccentricity this variability should be less than 1000, and in those cases where the runs did not reach more than a few thousand orbits we use either 500 or 100 orbit averages in Table \ref{table:exp}.

\begin{figure}[ht!]
\centering
\includegraphics[scale=0.3]{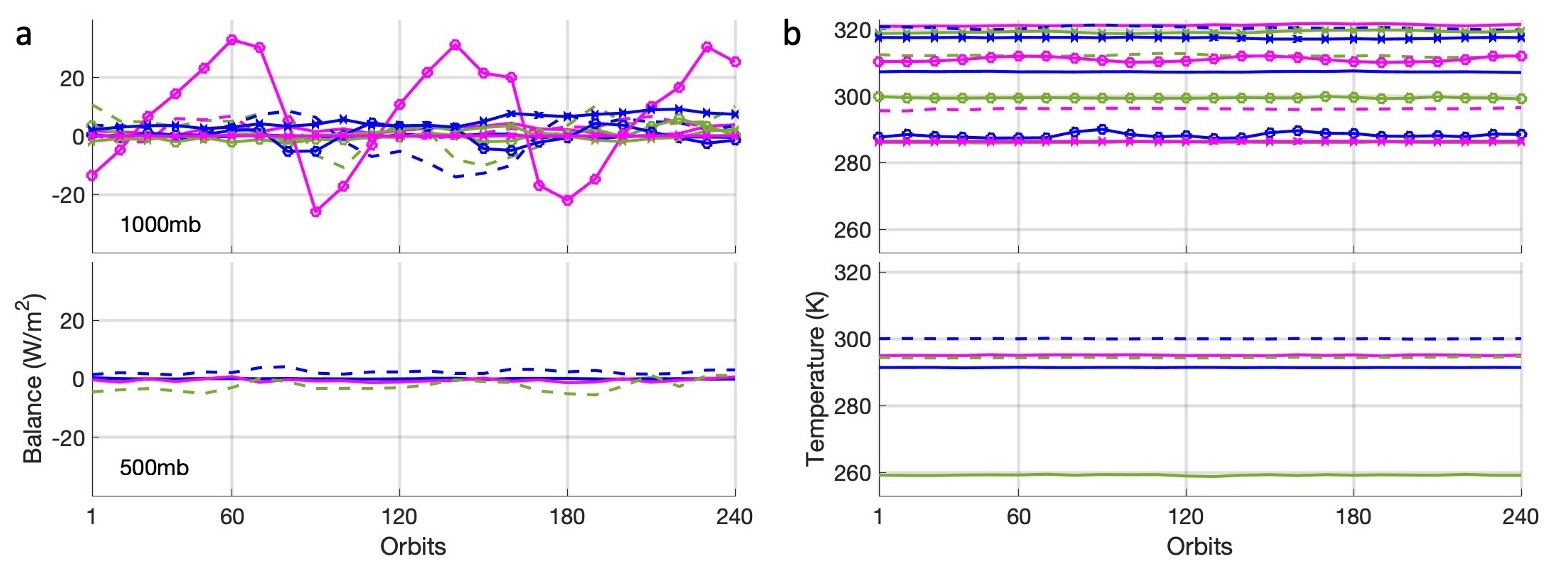}
\caption{This is the NRB at the TOA (A) and the mean global surface temperature (B) for the last 240 orbits of each simulation. The legend is the same as that used in Figure \ref{fig:1}. The large fluctuations seen in the magenta line with large circles (Sim 09 AP W2) correspond with the large imbalance seen in Table \ref{table:exp} for this experiment.}
\label{fig:nradtsurf}
\end{figure}

\end{appendix}
\newpage
\bibliographystyle{aasjournal}
\bibliography{references}
\end{document}